# Review: Artificial Intelligence for Liquid-Vapor Phase-Change Heat Transfer

*Youngjoon Suh[1], Aparna Chandramowlishwaran[1,2], Yoonjin Won[1,2]\**

[1]Department of Mechanical and Aerospace Engineering, University of California, Irvine, 5200 Engineering Hall, CA 92617-2700, USA
[2]Department of Engineering and Computer Science, University of California, Irvine, 4410 Engineering Hall, CA 92697-2700, USA

E-mail: won@uci.edu

Artificial intelligence (AI) is shifting the paradigm of two-phase heat transfer research. Recent innovations in AI and machine learning uniquely offer the potential for collecting new types of physically meaningful features that have not been addressed in the past, for making their insights available to other domains, and for solving for physical quantities based on first principles for phase-change thermofluidic systems. This review outlines core ideas of current AI technologies connected to thermal energy science to illustrate how they can be used to push the limit of our knowledge boundaries about boiling and condensation phenomena. AI technologies for meta-analysis, data extraction, and data stream analysis are described with their potential challenges, opportunities, and alternative approaches. Finally, we offer outlooks and perspectives regarding physics-centered machine learning, sustainable cyberinfrastructures, and multidisciplinary efforts that will help foster the growing trend of AI for phase-change heat and mass transfer.

Understanding phase-change heat transfer is important to numerous applications related to the advancement of energy conversion and thermal management systems. In particular, liquid-vapor phase-change processes continuously garner the interest of thermal scientists due to their ability to transfer large amounts of energy effectively.[1] Central to these processes is the nucleation of the dispersed phase (i.e., bubbles and droplets for boiling or condensation processes), where the continuous phases are liquid and vapor, respectively. The bubble or droplet dynamics can be manipulated by modulating experimental designs, including surface properties, structures, and materials, to push the limit of heat transfer performances.[2] Historically, bubble and droplet dynamics have been studied through the combination of nucleation theory,[3-6] thermodynamics,[7-10] and phenomenological correlations.[11] Yet despite phase-change heat transfer's century-long history, fully understanding its mechanistic relationships by linking experimental factors, complex nucleation statistics, and thermal performance remains an elusive challenge. A primary reason for this obscurity can be attributed to *data inconsistency* caused by a wide variety of operating conditions and experimental protocols along with measurement uncertainties.[12] Another reason is the *difficulty in quantifying boiling and condensation behaviors* because of their highly dynamic, complex, and high-dimensional nature.[13,14] Further difficulties arise in the *management and curation of data streams*, which have high implications for predicting and forecasting multi-phase flow patterns. The key concepts in phase-change heat transfer are listed in Box 1.

In the era of big data, transforming large quantities of data into useful knowledge plays an increasingly important role across various engineering disciplines. Artificial intelligence (AI), a term coined by John McCarthy in 1955, was defined by him as "the science and engineering of making intelligent machines." It has emerged as a dominant force for performing data-to-knowledge conversion and presents an attractive complement to traditional computational science. AI technologies simulate human intelligence processes through machines, particularly computer systems. Current AI technologies primarily fall in the category of "narrow" AI, which refers to an AI that can outperform humans on a particular task.[15] Machine learning (ML) is a subfield of AI where machines can learn without explicitly being programmed.[16] Neural networks trained using statistical methods make inferences from data that can lead to decision-making. Deep learning (DL), a class of ML inspired by our brain's neural network, can learn multi-level representations of data hierarchically.[17] While most existing AI solutions are still considered black boxes, the scope of ML and DL transcends mere nonlinear regression. Notably, artificial neural networks (ANNs) are now able to discover new material,[18,19] help model path- and history-dependent problems,[20] inversely design material structures,[21] and discover hidden physics from data,[22] as Box 2 describes the key concepts in AI and ML. Box 3 describes key concepts of digital inference that have seen great progress with advancements in AI and ML.

**Overview**

Within the phase-change heat transfer community, the plethora of recent AI-based approaches has motivated heat transfer researchers to explore their capabilities to advance fundamental nucleation sciences under a new paradigm (FIG. 1). Despite its potential, adaptations or discussions of AI technique have traditionally been slow in this field, which makes this untapped disruptive technology even more attractive. Fig. 2 summarizes the number of publications on AI and two-phase heat transfer research areas. According to the results, the publications for pure AI or two-phase heat transfer (approximately 100K) outweigh those from AI-integrated two-phase heat transfer research (i.e., AI + phase-change) categories by several orders of magnitude (approximately 100). While this ratio suggests that AI-integrated two-phase heat transfer research is relatively new, it also implies that this subfield is growing at a rate that exceeds that of both individual fields. Indeed, with the ongoing AI "gold rush" taking place in this field, the time is ripe to carefully review recent publications. To this end, the current article addresses the pressing need for a thorough review, explaining the core ideas of AI technologies, the innate challenges involved with phase-change processes and limited use of AI in the past (FIG. 1a), current implementation status and challenges (FIG. 1b), and future outlooks for two-phase heat transfer research (FIG. 1c). The article builds on prior reviews that introduced the intersection of AI and heat transfer in either a broad sense,[23] or for different applications facing disparate challenges than those of boiling and condensation heat transfer.[24-28] The critical review and summary will provide valuable insights into the implementation of AI for liquid-vapor phase-change heat transfer studies, with the goal to provide the thermal science community with a roadmap for future research.

**Fig. 1: Artificial intelligence (AI) in liquid-vapor phase-change heat transfer.**

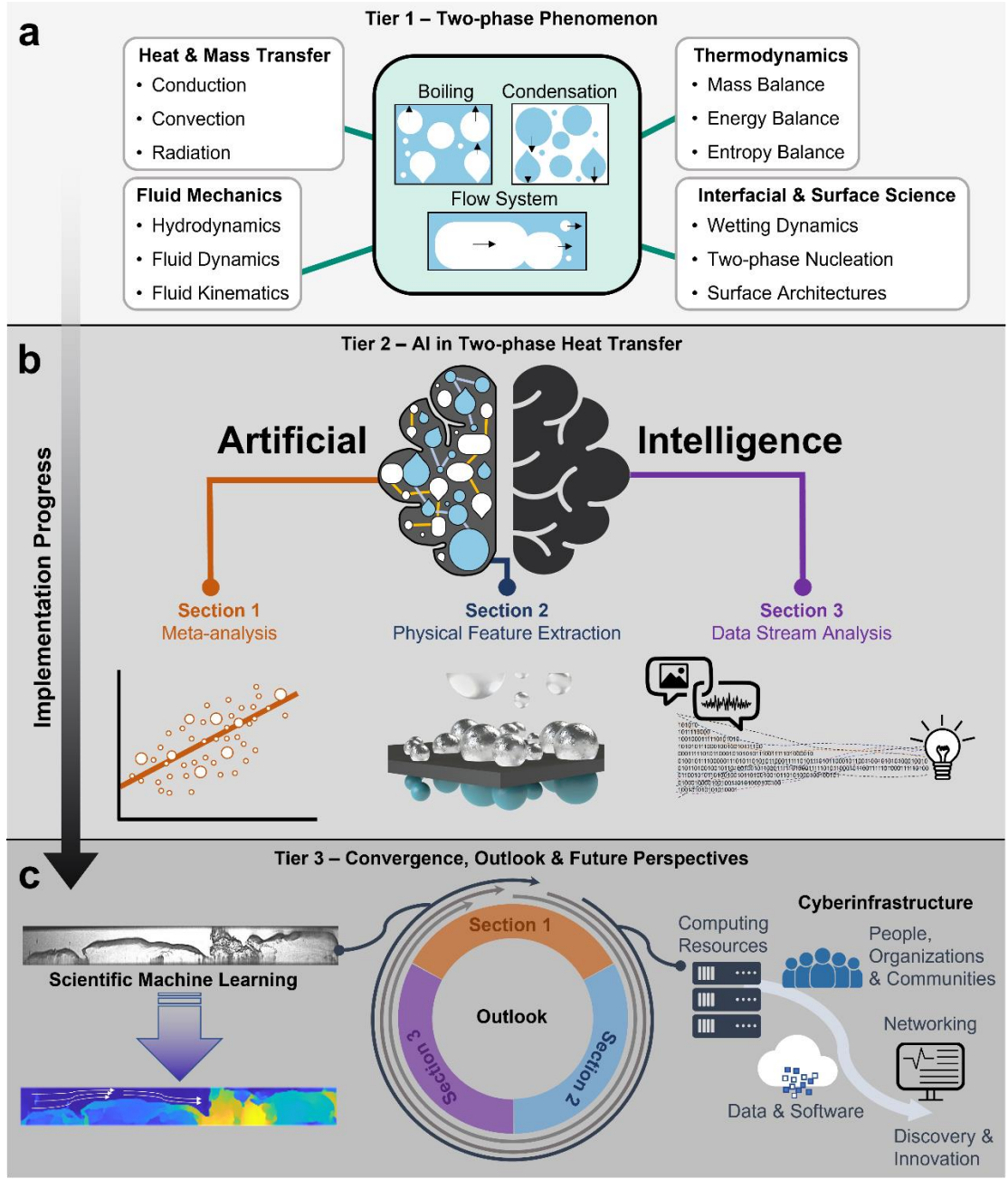

AI technologies offer diverse opportunities for scientific advances in phase change heat transfer. In this review, (a) the current challenges of phase-change phenomena are discussed along with (b) AI technologies categorized from an objective-based studies, followed by (c) their outlooks and future perspectives. The tiers are chronologically ordered to reflect the past, present, and future implementation progress of AI within this field.

**Fig. 2: Artificial intelligence and phase-change research publication trend.**

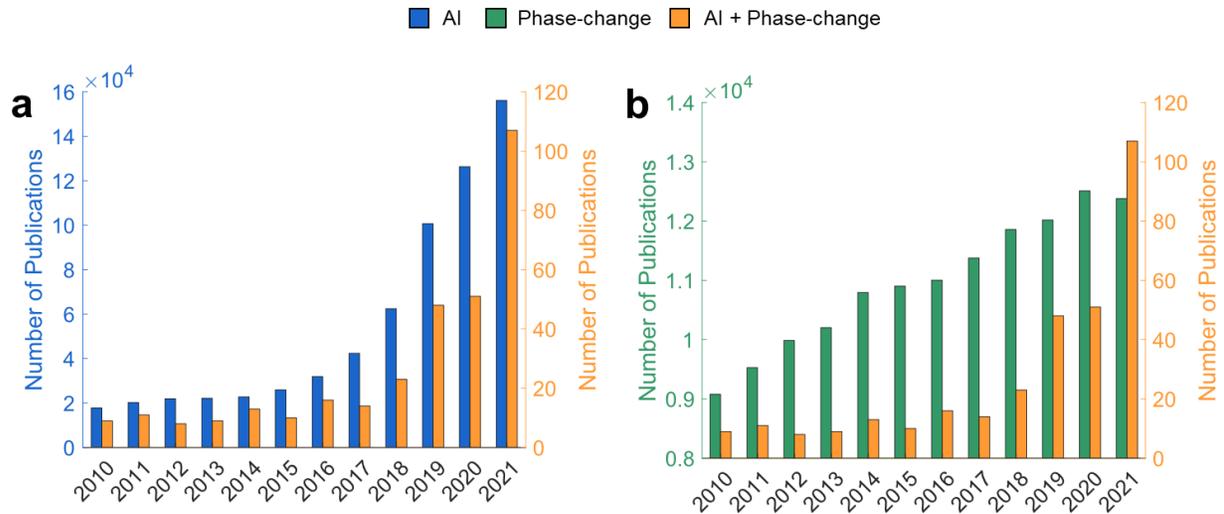

Graphs in (**a** and **b**) show the number of publications on AI and boiling/condensation research acquired from topic searches in Scopus on April 12, 2022, from years 2010 to 2021. The data under AI categories are obtained from keywords such as *artificial intelligence*, *machine learning*, or *deep learning*. The data for phase-change categories are found using keywords such as *boiling* or *condensation*. The data for phase-change studies involving AI (i.e., AI + phase-change) are obtained from the combination of the above two searches. In 2021, the number of publications on AI-based two-phase research increases by 365% compared to 2018 but still only occupies 0.8% of boiling and condensation research. The survey underlines the need for further studies that integrate AI into two-phase heat transfer research.

Box 1 | **Key concepts in phase-change heat transfer**

The boiling and condensation heat transfer performances are typically evaluated by measuring the heat flux in relation to temperature differences between the target surface (i.e., wall) and bulk liquid (i.e., superheat). We summarize important concepts for phase-change heat transfer below:

**Heat transfer curve.** The heat transfer curve showcases the temperature difference between the surface $T_s$ and the bulk fluid (usually at saturation temperature $T_{sat}$) and the corresponding heat flux (i.e., the rate of heat transferred per unit projected surface area), where the temperature differences are called the superheat and subcooling for boiling or condensation, respectively. Both boiling and condensation exhibit convection, nucleation-dominated, transition, and film-dominated regimes. The nucleation-dominated regime is commonly referred to as nucleate boiling and dropwise condensation regimes in literature and naturally lies at the heart of nucleation statistics-based research.

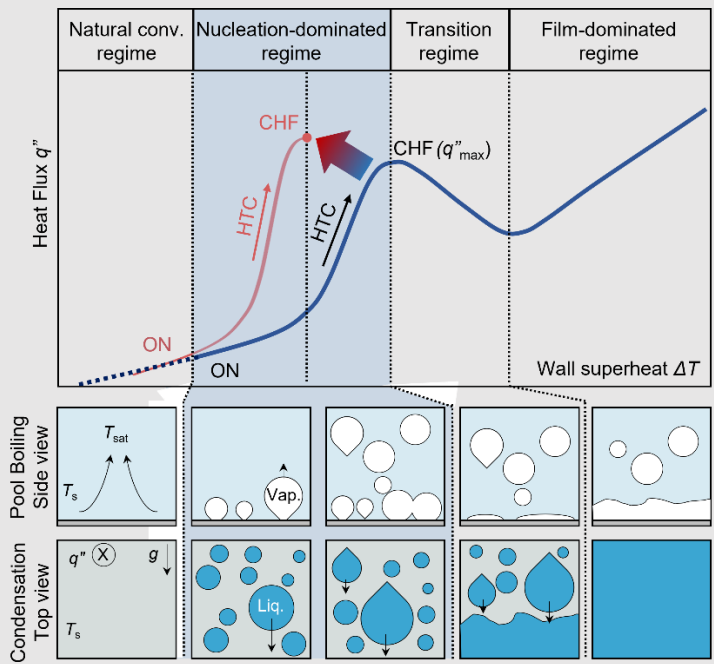

**Onset of nucleation.** The onset of nucleation (ON), also known as the nucleation incipience, is the point at which the dispersed phase first appears at the heat transfer surface.

**Heat transfer coefficient.** The heat transfer coefficient (HTC) is the proportionality constant between the heat flux and the temperature difference or the slope of the heat transfer curve. A higher HTC indicates that a greater amount of thermal energy can be transferred (i.e., absorbed or released) at a given temperature difference. In other words, the HTC quantifies the heat transfer efficiency of the phase-change process, and therefore a primary research thrust for thermal engineers is to increase the slope of the heat transfer curve. Because the HTC is intrinsically tethered to trade-offs between the desire for bubble or droplet nucleation and the necessity of removing them from the target surface, understanding the nucleation dynamics becomes an imperative factor to modulate the heat transfer curve.[29]

**Critical heat flux**. When plotting the heat transfer curve, the heat flux increases with increasing superheat or subcooling until a critical point. The corresponding heat flux to the critical point is defined as the critical heat flux (CHF), beyond which the HTC decreases sharply, and filmwise boiling or condensation modes develop.[29] Since film-dominated regimes show larger thermal resistance and less effective heat transfer performance, research has focused on strategies to remove bubbles or droplets from the surface before film-dominated regimes begin to dominate.[2, 29, 113]

Box 2 | Key concepts in AI and machine learning

Understanding AI for phase-change heat transfer requires familiarity with current technologies and their concepts. Below is a short overview of core terminologies in the broadest possible sense, without emphasis on mathematical detail.

**Artificial neural networks (ANNs)** are at the heart of deep learning algorithms. They are a hierarchical organization of neurons in layers containing an input layer, one or more hidden layers, and an output layer. The number of layers is the *depth* of the network. In practice, shallow ANNs may not be sufficiently expressive and have limited learnability.[30] Deep neural networks (DNNs), on the other hand, can learn more complex patterns at various levels of abstraction and are better at generalization.[31] ANNs are represented by a spectrum of different architectures.

**Multilayer perceptron.** A perceptron is the simplest form of an ANN binary classifier developed in the 1950s.[32] While a single-layer perceptron consists of only the input units and an output layer, a multilayer perceptron (MLP) contains one or more layers (i.e., hidden layers) between the input and output to learn more complex, nonlinear functions.[32]

**Convolutional neural networks (CNNs)** are a type of ANNs showing remarkable success in the detection, segmentation, and recognition of objects and regions in images.[33] CNNs typically comprise of convolutional layers for extracting feature maps, pooling layers to aggregate features, and fully connected layers for high-level reasoning prior to making predictions at the output layer.[31,34]

**Recurrent neural networks (RNNs)** are a class of ANNs that uses sequential or time series data. It includes a recurrent hidden state whose activation at each time depends on that of the previous time.[31] Hence, RNNs are capable of capturing and storing long-dependence relationships, which has led to their common use for temporal problems.[31]

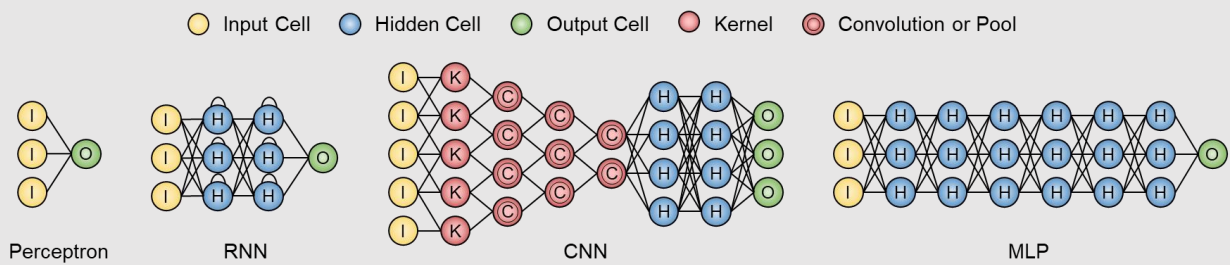

**Supervised learning** is a machine learning methodology where a model is trained using labeled data (meaning data that is annotated with correct answers or labels).[31] The training dataset plays a vital role for supervised learning algorithms to effectively learn. It is also worth noting that while supervised learning is currently the dominant learning paradigm in machine learning, it is not representative of human and animal learning.[35]

**Unsupervised learning**, in contrast to supervised learning, focuses on discovering underlying patterns in unlabeled datasets.[33] Although unsupervised learning models learn from raw data without any predefined labels or target outputs, there is often a need for larger training data to yield desired results as complex phenomena can often be challenging to decode without domain expertise or prior knowledge.

**Transfer learning** is the adapting of a pre-trained model for a wide range of downstream tasks, using minimal additional data for each new task.[36] For example, leveraging feature representations from models already trained on large datasets such as ImageNet[37] and Common Objects in Context (COCO) datasets,[38] have become a central part of modern object detection models.[39]

Box 3 | **Key concepts of digital inference**

Digital inference has assisted thermal engineers over the last several decades. With the advancement of DL models for CV, researchers can now extract nucleation features with unprecedented resolutions.

**Computer vision.** Computer vision is a field of study that focuses on enabling computers to infer meaningful information from images and videos.

**Image classification.** The origin of digital image inference can be traced back to automated image classification approaches, where predictions describing the image or a list of classes of objects in an image are provided based on their classification scores.[40]

**Object detection and localization.** While image classification by itself has merits and practicality, the features that it uses have limited capability to provide physically meaningful descriptors for nucleation processes. The next incremental step from coarse to fine image inference is object detection and localization, the task of determining where objects (i.e., bubbles and droplets) are located in a given image (object localization) and which category each object belongs to (object classification).[41] At this stage, valuable information such as the rough size, distribution, and spatial coordinates of the nuclei can be collected in the form of bounding boxes or centroids.

**Semantic segmentation.** Semantic segmentation allows for more detailed spatial analysis (e.g., void fraction) by labeling every pixel according to its class.[42]

**Instance segmentation.** Progressing for finer inference leads to instance segmentation, which provides separate labels for each object belonging to the same object class.

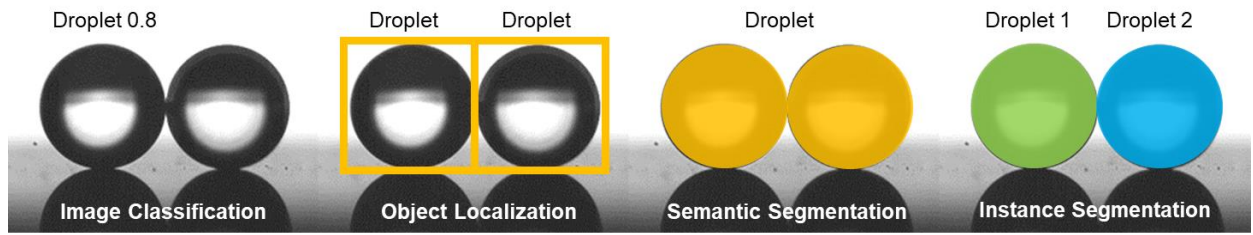

**Object tracking.** Object tracking for collecting high-fidelity spatio-temporal nucleation statistics is a hierarchical process that has important implications for better understanding mechanistic two-phase nucleation principles. At the lower end of the hierarchy, spatial features such as the two-phase demography (i.e., size, location, distribution) or void fraction provide foundational information describing how bubbles or droplets nucleate, coalesce, and depart. At the higher end of the hierarchy, the extracted features are linked to time, rendering features (i.e., growth rate, nucleation interaction type, departure frequency, departure size, rising velocity) that can express the surface activity with much more context.

## State-of-the-art AI technologies for phase-change heat transfer

The article first discusses current AI technologies available in the heat transfer community (FIG. 1b) categorized based on how they target major and relevant two-phase heat transfer problems existing today. AI technologies are broadly divided into three categories, namely *meta-analysis*, *physical feature extraction*, and *data stream analysis*.

### Meta-analysis

Meta-analysis on multiple datasets is generally conducted to provide a holistic description of phase-change heat transfer by either corroborating consistent datasets or revisiting conflicting datasets, usually in the form of tabular data.[43] This is important because both experimental and computational data collection in this field is expensive and requires significant upfront investment. Additionally, various combinations of experimental factors—including the material, thickness, surface roughness or structures, and surface/working fluid wettability—exert significant influence on the datasets.[29] Moreover, the cumulative data available today have high variance and is hypersensitive towards various operating conditions,[44] mounting protocols,[45] characterization procedures,[46] and the personnel's experience level. Once meta-datasets are collected, data dimensionality is orders of magnitude larger, making it challenging to understand. To address these challenges, a recent movement in the thermal science field seeks to exploit available scientific advances through *machine learning-assisted meta-analysis* (FIG. 3).

*Data-driven hypothesis approach*
ML models can be efficient by assisting professionals to make holistic data-driven hypothesis development by learning big data. ML-assisted meta-analysis can play a role in learning knowledge from data and extracting data from knowledge, as described in FIG. 3. The first step introduces building hypotheses based on accumulated datasets to answer scientific questions in phase-change heat transfer (FIG. 3a). After two-phase experiments are designed and performed (FIG. 3b), the experimental factors (tabular data) (FIG. 3c) are then used to train neural networks to find the relations between the experimental factors and their output (heat transfer performance) (FIG. 3d), thus helping researchers inform their understanding and decisions with data.

**Fig. 3: Holistic data-driven workflow.**

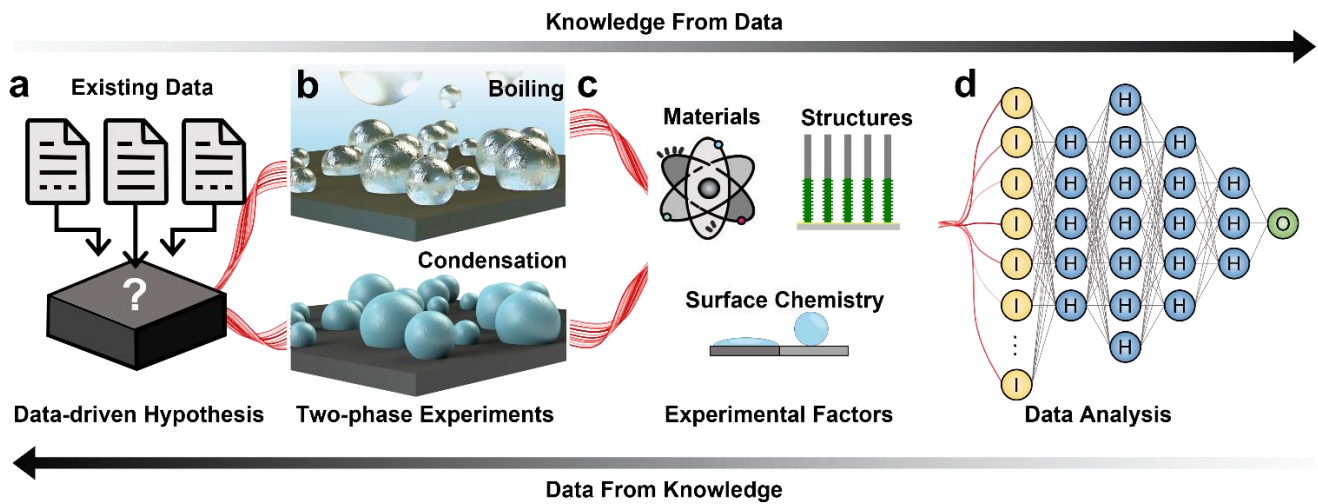

The cycle consists of **a** | data-driven hypothesis, followed by **b** | two-phase experiments. **c** | Experimental factors are then collected to build **d** | artificial neural networks for meta-regression analysis.

Unlike traditional theory-driven approaches, which rely on the knowledge of the underlying mechanisms of the observed phenomena, data-driven approaches can uncover statistical relationships and patterns solely from data even without explicit knowledge of the underlying physics or mathematical functions. This characteristic enables ML models achieving high-accuracy predictions to be trained in a matter of hours compared to the extensive period and resources required to develop new heat transfer theories. However, data-driven approaches typically require larger diverse datasets for generalization, the size of which depends on the specific problem, and lack the interpretability offered by theory-driven approaches.

*Regression models*

Most ML-based meta-analysis models utilize regression analysis to link experimental factors with non-dimensional numbers or global heat transfer quantities such as CHF, HTC, or pressure drop over the entire system (Table S1). The generic feed-forward ANN architecture, also known as a multi-layer perceptron (MLP),[47] and random forest (RF) methods have been the most popular,[47] probably because they are some of the most well-understood techniques in literature today. Both ANNs (or MLPs) and RF are exceptional at modeling complex nonlinear relationships, with the general idea of approaching problems by deconstructing them into smaller simpler units. ANNs are organized into layers of interconnected nodes, in which "weights" are assigned to represent the value of information assigned to an individual node.[33] Similarly, RFs statistically identify important features, create multiple randomly chosen weak decision trees, and collect their votes to make the final electoral decision. To handle complexity and overfitting issues, ANNs can adjust hyperparameters such as the number of hidden layers and units, regularization techniques, and learning rates, whereas RFs can modulate the number of trees, the maximal size or depth of the single tree, and sampling rate.[48] ANNs typically have more flexibility in their architecture designs, while RFs, although less explainable than conventional decision trees,[49] are reported to possess potentially better interpretability than the so-called black-box predictions of typical ANNs.[48] In general, no single ML model works the best across all problems, and therefore it is good practice to compare the performance of different techniques or create a hybrid model that combines multiple feature types.[50] Other ML algorithms that have previously been investigated include the support vector machines (SVMs),[47] boosting algorithms,[47,51] cascade feedforward (CF) networks,[51] radial basis function (RDF),[51] adaptive neuro-fuzzy inference system (ANFIS),[52] deep belief networks (DBNs),[53] convolutional neural networks (CNNs),[54] and physics-informed neural networks (PINNs).[55,56]

*Challenges and opportunities*

The training data for meta-analysis models are often inadequate or sporadic, which can pose a great challenge for developing robust ML models. The training data from experiments that exist today are sparse due to the prohibitive cost of acquiring more experimental data points, especially considering the difficulties associated with replicating most nm – μm scale bottom-up fabricated surfaces.[2] On the other hand, training data from simulation models are relatively denser, but are limited to extremely simple cases. Recent studies have suggested that *adding physics into the learning process* and *developing surrogate ML models* can address these issues by efficiently training ML models with limited training resources.[57,58]

First, breakthroughs in physics-informed machine learning (e.g., PINNs) have demonstrated that data from both measurements and simulations, and the underlying laws of physics can be integrated into the loss function of the neural network, thereby providing strong theoretical constraints to prevent models from generating unreasonable predictions. This not only allows models to train with less data than is typical for over-parameterized DL models by narrowing the search space, but also enables them to estimate values that extend beyond the trained data scope. Although a major challenge remains in the search and development of physical models that best represent the increased volume, velocity, and variety of available data, other communities suggest that PINNs can succeed,[22,59] even for such ill-conditioned problems, in cases where other models cannot. This suggests new research opportunities for meta-analysis implementations in the phase-change heat transfer community.

Second, developing surrogate ML models that sensibly augment the training data to generate diverse problem instances can improve the model's generalizability even with minimal training data. A surrogate model is a trained emulator that learns to approximate solutions through known input-output behaviors when an outcome of interest cannot be easily measured or computed. Therefore, filling in the gaps of these scattered data points will allow the models to learn higher-dimensional insights that can be extended to unexplored study regimes.

**Physical feature extraction**

Beyond the metadata analysis, it is imperative to extract physically meaningful features from visual data. Digital inference and extraction of datasets have the potential to enhance our understanding of new physics within large datasets by enabling a full description about two-phase physics. One area of particular interest is the quantification of nucleation behaviors during two-phase processes, which has been a focus since the pioneering work of Nukiyama and Schmidt.[60,61] While visualizing nucleation behaviors has naturally become an inseparable part of phase-change studies,[62-67] the sheer complexity and volume of bubble and droplet activities (FIG. 4a and c) make quantifying these activities a daunting challenge. In this regard, AI-assisted CV can be effective at performing image analysis tasks with high-level accuracy but with far greater bandwidth.[13,14,68] Most noticeable advances come from characterizing two-phase nucleation features with the basis of modern CV tasks from CNN models built for digital image inference tasks (Box 3).

*Overview of nucleation statistics and heat and mass transfer*

Concurrent with the advances in AI and CV has been an increasing demand to support a stronger connection between bubble and droplet statistics and thermal performances. For example, a significant portion of extant bubble dynamics studies only reports low- to moderate-heat flux regimes, leaving an extensive range of high-heat flux regimes unexplored during pool boiling (FIG. 4b; Table S4). Similarly, droplet nucleation statistics are also highly indicative of hidden condensation heat and mass transport mechanics. However, only a handful of studies have attempted to extract spatio-temporal instances (STI) that are essential for decrypting heat transfer analysis. According to a brief literature review of 56 recent studies reporting droplet dynamics from 2015 (FIG. 4d; Table S5), the vast majority report only spatial features or spatio-temporal group-based (STG) analysis, where relatively simple traditional CV algorithms or manual labor are still favored (FIG. 4d). In addition, only 11% of the reviewed studies demonstrate heat and mass transfer analysis coupled with droplet statistics, further underlining the need for a stronger connection between the two approaches (FIG. 4d inset).

*Traditional computer vision*

For a very long time, rudimentary image processing algorithms have extracted basic nucleation features to describe phase-change phenomena.[69-72] These traditional CV approaches use relatively simple processing algorithms such as binarization, denoising, flooding, and interactive separation of connected objects to capture the most basic physical descriptors such as bubble or droplet size and distribution.[70,73] While these approaches are quite useful and swift, they still ultimately rely on handcrafted features that inevitably require significant setting optimizations often under-reported in the literature.[74,75] To address this issue, there have been efforts to develop shape filters or reconstruction algorithms such as built-in circle finder scripts or multi-step shape reconstruction procedures that specifically exploit the circular shape of bubbles and droplets.[73,76-80] Yet, due to these methods' inherent imposition of rules, it has been difficult for traditional CV models to adapt to image inventories from the broader research community, with numerous studies reporting unstable predictions, even with a slight change of lighting.[79,80]

*Machine learning-assisted computer vision*
Recent integrations of DL models into CV have brought a paradigm shift in how researchers perceive and utilize visual data to tackle difficult problems. DL models "learn" salient features of the target object, thereby requiring less expert analysis and parametric fine-turning, and possess superior flexibility in adapting to custom datasets compared to traditional CV approaches.[81] For the time being, studies implementing models with CNN backbones have seen the most success due to their proficiency at handling image data.[13,14,68,82]

For boiling, the ability to extract spatio-temporal bubble statistics has proven direct implications for heat flux partitioning analysis (See Supplementary information Box S1), the process of understanding what heat transfer mechanisms constitute the boiling heat flux at any given stage,[83-85] as well as flow instability analysis.[119] However, extracting a sufficient amount of classical boiling features (i.e., departure diameter, nucleation site density, departure frequency) from imaging data has been an inherently burdensome and evasive task. The

**Fig. 4: Overview of nucleation dynamics during phase-change processes.**

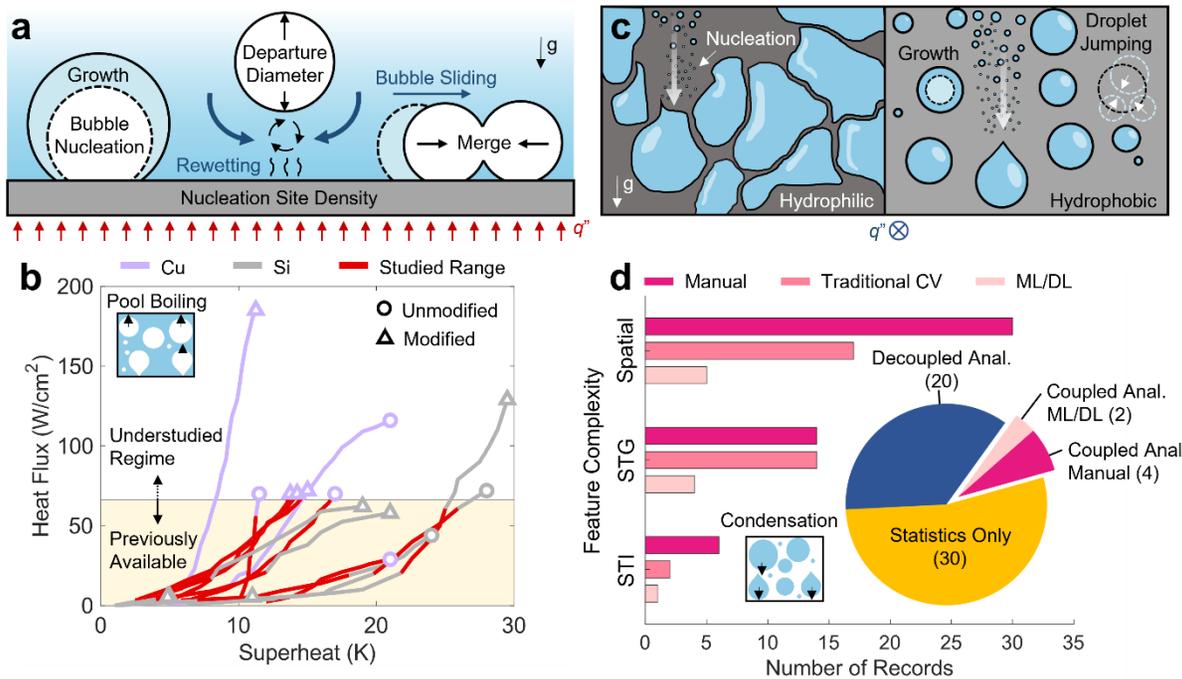

Quantifying the nucleation dynamics is crucial to better understand two-phase phenomena because they can explain the heat transfer performance of **a and b |** boiling and **c and d |** condensation. Literature review of **b |** pool boiling reveals that a significant portion of the heat transfer curve is understudied in terms of bubble dynamics. The only studies that have been able to investigate nucleation dynamics across a relatively wide heat flux range are ones that utilize bottom-to-top infrared (IR) imaging techniques, in which the translation from bottom IR bubble statistics to actual bubble morphologies are not well understood. Literature review of **d |** condensation shows that 86% studies either excluded heat transfer characterizations (50%) or measured heat and mass transfer in a decoupled manner, independent of droplet statistics. Only 11% of the reviewed studies demonstrated fully coupled heat and mass transfer analysis based on droplet statistics, where 4 of the 6 coupled analysis studies were done manually. The references that have helped form the conclusions in this figure for pool boiling and condensation are provided in Table S4 and Table S5, respectively.

autonomous curation of high-quality bubble statistics at the large scale has been shown to be achievable through DL-assisted segmentation and tracking,[118,119] enabling the quantitative mapping of bubble dynamics between different boiling surfaces at unprecedented resolutions.[13]

Condensation studies have shown that heat transfer can be effectively quantified by collecting statistics of group of droplets or individual droplets (See Supplementary Information Box S2) and integrating them over the entire surface.[14,82] A study on group-based droplet statistics has shown that the overall time-averaged droplet shedding frequency can be used to calculate the heat transfer rate by using Eq. S5 and S6, which ultimately reduced the high (20 – 100%) measurement uncertainty of conventional temperature and flow sensors less than 10%.[82] In another study, collecting single-droplet-level statistics imposed with energy balance equation has allowed for the mapping of the evolving heat transfer of condensing surfaces with extreme spatio- and temporal- resolutions of 300 nm and 200 ms, respectively.[14]

The recent demonstrations of statistics-driven heat transfer quantifications are particularly exciting because they enable the robust comparison and de-coupling of the multi-dimensional relationship among nucleation parameters, heat transfer performances, and structural design. In addition, these techniques can be used to simultaneously compare the heat transfer performances on a single surface with different structures or wettabilities,[82] which not only minimizes time and labor resources but also essentially eliminates uncertainties caused during separate experiment trials.

A summary of studies that quantify boiling and condensation nucleation behaviors from 2015 - 2022 are provided in FIG. 5 and Table S4 – S6. We define real-time prediction for models that have the capability of making predictions in live streams, quasi-real-time prediction for models that require an interventional processing step, and extensive prediction for models that report an undefined amount of time. The extracted physics are categorized as spatial, STG, STI, and hidden, which refers to the non-intuitive features of the hidden layers in ANNs.

*Challenges and opportunities*

DL-based nucleation feature extraction models are still far from being widely adopted in the heat transfer community, as evident in FIG. 5. The fact that despite modern advances in CV, manual analysis is still a highly favored method for quantifying nucleation dynamics suggests that there exists a gap between thermal and computer scientists. Therefore, further efforts should be considered to make these ML-based quantification methods more efficient and user-friendly by *fine-tuning the models to be suitable for specific two-phase conditions* and *improving training methodologies.*

One challenge is about how two-dimensional image sequences can accurately represent three-dimensional phenomena. This representation often results in different subproblems specific to two-phase phenomena such as occlusion (situations when an object fully or partially hides another object of the same class from the optical viewpoint), cluster, and low resolution. One method that researchers have investigated involves the refining of initially predicted results through multi-step shape reconstruction processes.[92-95] While initial object detection methods may vary (i.e., bounding box, regional proposal, image segmentation), the concept of reconstructing the digital inference from initially rough object outlines to its true shape remains the same.[92-95] However, these models are not programmed or trained to inherently recognize the occlusion phenomena themselves and are therefore limited to weakly occluded instances. Researchers soon realized that the model's predictive capability can be finetuned to specific problems by modulating the training dataset. By customarily labeling droplets that were blurred by motion, researchers successfully trained the model to predict images captured at low framerates.[82] In another study, a model was trained to accurately predict occluded droplets by custom-labeling droplets that formed underneath larger droplets.[96] Other efforts include the use of synthetic datasets created from generative adversarial networks (GAN),[92] or simulations.[97] The rationale of using such synthetic datasets to train is to carefully design a dataset with a controlled amount of object occlusion, nucleation sites, object

**Fig. 5: Summary of processing time versus extractable physics.**

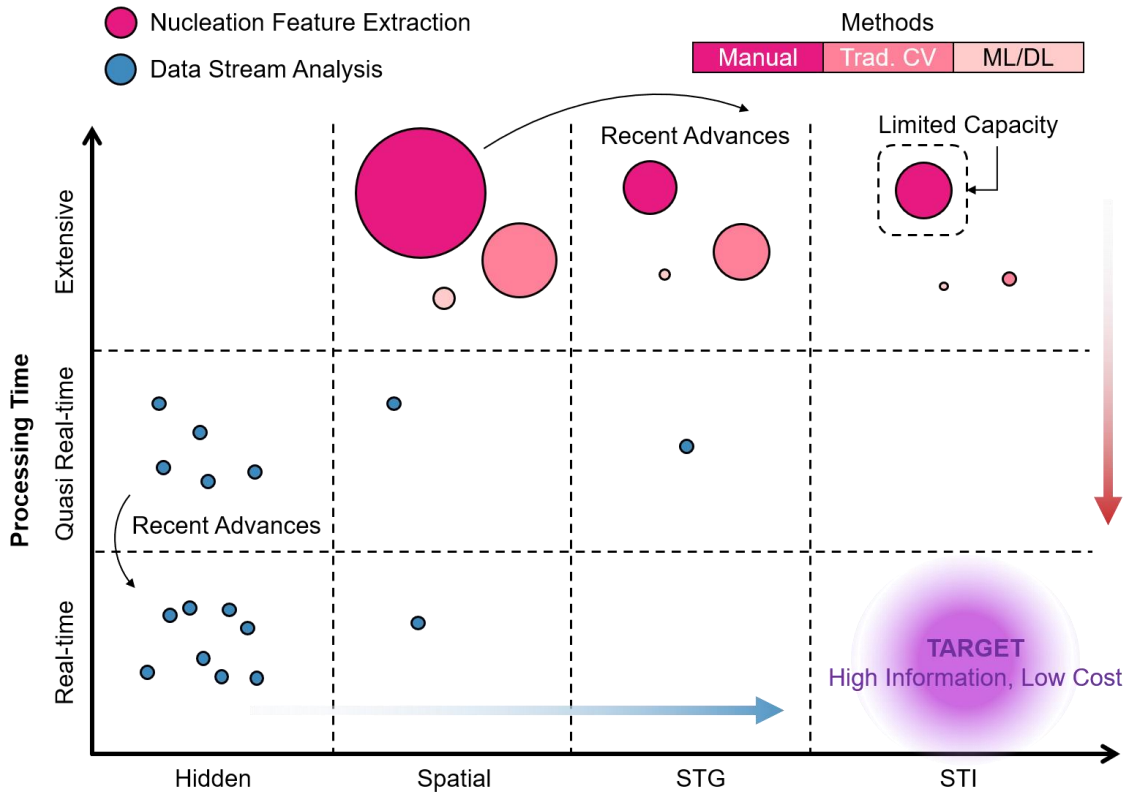

The regime map reveals a tradeoff between the processing time and the extent of physically meaningful features extracted. A summary of 94 studies that reported bubble or droplet dynamics is shown in red tones, where the color indicates the methods utilized to extract nucleation features, and the circle size represents the number of studies normalized by the total number of studies. Individual data stream analysis studies are listed in Table S6.

density, and size to emphasize model learning for difficult scenarios. Hence, there must be a continuous effort to explore various training sets to investigate what and how the model learns, depending on the research objective.

Another challenge is centered upon data preparation. Due to the intrinsically complex phenomena being studied, data labeling requires experts. Most upcoming models use supervised learning, which requires training datasets in the form of labeled images. The image labeling process, called image annotation, can take up to several months to manually prepare a complete dataset for training and validation.[14] More often, the ground truth for the labeling is subjective and involves human intervention, making unanimous agreement among annotators impossible.[86] To combat such discrepancies, researchers have proposed using cost-effective, human-in-the-loop curations or semi-supervised learning that can alleviate annotation labor and increase labeling consistency.[87,88] For example, a recent study proposed a multi-step cycle of initial pretrained model predictions, human-aided error identification and correction, and retraining on an updated dataset.[88] While training an initial model to start the loop remains an ongoing challenge, abundant research opportunities exist developing in fully autonomous labeling pipelines using deep generative models to conditionally generate sequential visual scenes that mimic real states,[89-91] or even label-free unsupervised learning-based nucleation feature extractors to reduce time and labor involved in training and validating models.

Above all, the effort-to-reward ratio for exploring key features that describe underlying boiling and condensation mechanisms significantly favors researchers today. For example, features that connect nuclei interactions and heat transfer have been rarely discussed in the literature but can now be studied in depth with light algorithmic adjustments. Furthermore, big data trends of long-term nuclei statistics can provide rational mechanistic insights but are yet to be reported. By adding a tracking module to an object detection framework, researchers have proved that features with substantially higher-order insights could be extracted from videos with minimum modifications.

**Data stream analysis**

Another theme in AI research related to phase-change is to comprehend the *transient* nature of metadata or visual data during phase-change processes. Given the strong yet complex correlation between "in-motion" data and heat transfer performance, researchers have classified the multi-phase flow patterns into various categories.[2,29] For example, the cyclic nucleation behaviors that constitute the heat transfer curves (Box 1) have traditionally been categorized into natural convection, nucleation-dominated, transition, and film-dominated regimes.[29] Evolving flow patterns in flow boiling systems have been classified into liquid, bubbly, slug, annular, mist, and vapor regimes.[98] Real-time data stream analysis using AI has the potential to identify or predict nucleation phases and patterns, which can be applied in smart boiling and condensation systems (FIG. 6a) for adaptive or predictive decision-making.

*Discrete models*

It is important to identify transient and unstable two-phase patterns into discrete class outputs, which is equivalent to the traditional ML classification problem.[99] The successful identification of patterns will help devices or systems operate under optimal conditions depending on their specific requirements, as illustrated in FIG. 6. For example, nucleate boiling in passive two-phase heat transfer devices, such as heat pipes, is typically detrimental to the device's performance due to significant heat and mass transfer reduction caused by entrapped vapor bubbles inside the wick.[100] By contrast, the creation of small and replenishable droplets and bubbles is favorable to facilitating heat transfer in high-power heat exchangers.[1] In these nucleation-favoring applications, identifying the transition from the nucleation-dominated regime to the film-dominated regime is crucial to maintain high heat transfer performances as well as safe operating conditions.[1,2]

Despite its significance, there have been only a few ML studies that demonstrate these classification tasks (FIG. 6b). The classification can be executed by using different types of data, ranging from signal patterns collected from sensors to qualitative visual descriptions of gas and liquid phase morphologies.[101] Since visual descriptors are outwardly more intuitive, the majority of studies have utilized visualization-based ML classification models (Table S6). Naturally, CNN models have been heavily adopted to stream classification problems, with many models exceeding 98% prediction accuracy.[68,102] One study showed that CNN models could even deal with difficult visual cases, such as the transition from nucleate to film boiling, which trained experts cannot distinguish with any reasonable certainty.[68] It is worth noting that approaches that use structured data (i.e., non-imaging data) are also important, for they allow data stream classification where visualization might be limited or even inaccessible (e.g., in-tube flows) but is rarely addressed.[103,104]

*Continuous models*

In addition to discrete outputs, ML can also be employed to predict and forecast continuous outputs (FIG. 6c) using regression analysis. Regression methods are proficient at determining casual relations between independent and dependent variables and therefore have major applications for numerical and visual time-series forecasting (FIG. 6c; Supplementary information Box S3).[105] Both time-series forecasting methods utilize transient data streams to predict future outcome trends (FIG. 6c), but visual forecasting approaches reconstruct the data into visual scenes and are viewed to be more computationally expensive.

Continuous models are especially appealing for prediction and forecasting tasks because they can potentially be trained to encode the context of the two-phase phenomena, offering a great advantage over traditional sensors and probes. The models often hypothesize that a DL model can be trained to predict heat and mass transfer using secondary, higher-resolution characterization methods, such as visualization as an input. This would effectively allow the model to make more "physically-intuitive" or "scene-incorporating" predictions. Recent studies have confirmed this hypothesis, showing that vision-based models are highly sensitive to physically interpretable features such as the number and size of bubbles,[102] and there are strong correlations between collected bubble parameters and boiling heat flux intensity.[68]

*Challenges and opportunities*

Although the main keywords of data stream analysis are "real-time" and "forecasting," neither one of these goals has been adequately met in the existing literature. A noticeable flaw is the use of previously collected data streams instead of live-streaming data. Therefore, future goals must shift the current paradigm of data stream analysis to *streaming data analysis*, where models train on more realistic data. The main keys to achieving this goal are to improve model transferability, explore indicative features, and expedite prediction time.

The first key is to improve model transferability factoring in the diverse boundary conditions that contribute to the complex and heterogeneous two-phase behaviors. While data supports that models can predict accurately settings familiar to the environments in which they were trained, studies customarily mention the need for additional training datasets for new boundary conditions.[68,102] Data augmentation or transfer learning techniques can be leveraged to remediate this issue, but only to some extent because they instigate the constant involvement of computer engineers to update the model.[68,102] Considering the inevitability and necessity of building *foundational models*, it remains an open question as to whether sufficient data can be collected and used to train models that incorporate the full complexity of the heterogeneous boiling and condensation phenomenon in varying conditions.

Another method to improve ML models is to identify *indicative features* that best represent the two-phase time-series problems. The features can be identified by engineering new features or selecting a new set of features.[106] As such, this process requires significant domain expertise in two-phase heat transfer to understand new features that can capture complex and nonlinear heat transfer performances during phase-change. Due to these aspects, unraveling the enigmatic features that can realize real-time prediction and forecasting is an ongoing challenge. For instance, a recent study utilized dimensionally-reduced image data to forecast future bubble morphologies based on a bidirectional LSTM network.[107] Images provide a visual representation of nucleation dynamics, which ultimately reflect the influence of experimental factors on the system. Consequently, investigating generalizable features that establish connections between nucleation patterns and desired outputs, researchers can potentially achieve a deeper understanding of nucleation phenomena and develop models that can be applied across various experimental conditions, leading to broader and more robust scientific insights. Future work should improve forecasting ML models by exploring a wide range of data types, including structured data (i.e., numerical data, or categorical data; low-dimensional), unstructured data (text, images, audio, video, or multi-modal; high-dimensional), and their hybrids (i.e., multimodal deep learning).

Additionally, it is crucial to evaluate the features based on their processing time, ensuring that they enable real-time prediction and forecasting. Models that utilize structured data inputs have quicker prediction speeds but at the expense of lower levels of physics. In contrast, models using unstructured data can extract physically meaningful features but are time-consuming, as showcased in FIG. 6. Typically, graphic datasets are highly information-dense and therefore can easily be inundated with redundant or irrelevant pixels (i.e., background data or noise), which requires additional data processing steps to reduce the data dimensionality or filtering

**Fig. 6: Data stream analysis.**

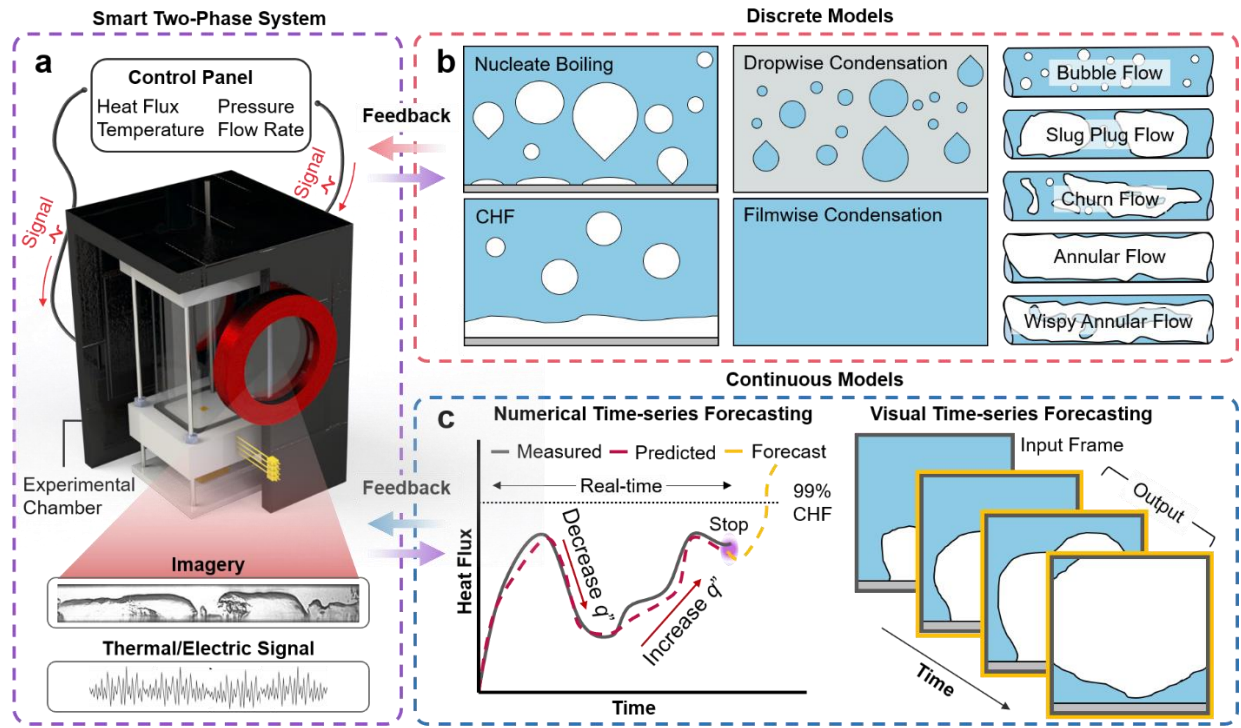

**Understanding streaming data can be utilized to build a | smart two-phase systems that make real-time adaptations by receiving** feedback from systems using either discrete or continuous models. **b |** On the one hand, unstable and transient multi-phase flow patterns are classified into discrete outputs when using discrete models. These classifications can offer instant feedback to change device settings when differences in flow patterns are detected. **c |** Continuous models have implications for real-time predictions as well as time-series forecasting. The models can use time-series texts or visual images to forecast a CHF event and stops the system in real-time before the event occurs.

important information only.[108] While dimensionality reduction methods (e.g., principal component analysis, resolution scaling) partially address the latency issues of vision-based models, they do so by sacrificing high-frequency image features and by introducing additional preprocessing steps, which have been underreported in many cases. An alternative solution is to selectively exploit only the important information. One recent study showed that this could be done by using neuromorphic event cameras that emulate the human retina, thereby only outputting "meaningful events" (i.e., pixels that display changes in brightness).[108,109] The fundamentally different way that event cameras encode visual scenes allows them to be proficient for applications with strong constraints on latency, power consumption, or bandwidth.[110] In addition, the ability to remove unwanted, stationary background noise potentially allows neuromorphic cameras to resist from generalizability issues stemming from measurements across multiple labs and institutes which may use different experimental setups. While the practical employment of event cameras is vastly underexplored, models that classified boiling regimes by using simulated event streams demonstrated a 300% increase in prediction speed compared to ML models reported elsewhere. Along the same lines, other potential state-of-the-art DL models, such as vision transformers and spiking neural networks, may find applications in this domain for real-time predictions and forecasting.

**Outlook**

The rapid growth of AI-based solutions has highlighted the potential benefits for the heat transfer community. The impact of these technologies will depend on the degree of active exploration by researchers into these exciting advancements in computer science. Moreover, it should be emphasized that the AI technologies discussed in this review are not independent attempts, but rather, share a symbiotic relationship (FIG. 1c). Meta-regression analysis allows for holistic decision-making to find optimal two-phase heat and mass transfer performances. Physical feature extraction generates big data that has not been available in the past, which can advance our understanding and serve as the basis for new physical models. The knowledge gained from these attempts is instilled into data stream analysis models to demonstrate futuristic smart phase-change systems.

**Future Perspectives**

*Physics-centered machine learning*

Quantifying the turbulent and spontaneous characteristics of the multi-phase physics involved with phase-change heat transfer processes with adequate quality that meets the demands of modern research has always posed a great challenge to researchers. Physical sensors or probes, which have been the gold standard for quantitative heat transfer characterizations so far, have limited spatial resolutions due to their genetic size. Besides that, optical measurement techniques such as infrared (IR) thermometry, fluorescence thermometry, and particle image velocimetry are limited to small domains and laboratory settings. On top of all this, there is no existing tool that can measure fluid and vapor properties simultaneously, which is crucial to understanding the underlying transport physics between the two phases. Although high-fidelity two-phase simulations have evolved over the past decades as an alternative solution to this problem, they are computationally expensive and often intractable without large-scale clusters or supercomputers.

The challenges associated with acquiring high-quality data might be addressed by leveraging AI or ML models that combine data with domain knowledge. Advances in ML models that incorporate known physics have demonstrated great potential to achieve accurate and interpretable results by satisfying both the observed data as well as the underlying physical laws.[114-117,120] In the context of phase-change heat transfer, the physics that domain experts can incorporate into ML models lies on such a wide spectrum, ranging from basic continuity equations to instability laws, that there exists abundant opportunities for researchers to explore integrating physics into their ML frameworks. An example of such implementations could be developing ML models that can use experimentally measured properties of one phase (e.g., vapor or liquid) and predict the velocity, pressure, and temperature fields of the other, which is the crux of many two-phase heat transfer problems. In such examples, the physics-respecting properties of ML models will allow domain experts to assess the model's validity and reliability in a more interpretable manner when it extrapolates beyond the range of observed data. Therefore, we envision that the fusion of data-driven models and domain expertise presents a promising avenue for accelerating scientific discoveries by combining the advantages of both approaches.

*Building cyberinfrastructures for two-phase heat transfer*

Beyond answering core science questions, it is imperative to develop an array of cyberinfrastructure (CI) technologies (FIG. 1c), including open-source and reusable data and scalable algorithms and software for long-term sustainability. The envisioned CI ecosystem should be agile and integrated to catalyze new transformative discoveries in heat transfer. This necessitates sustained community collaboration and continuous improvement at every stage.

Due to the nature of nonlinear, multi-dimensional, and multi-modal features during phase-change processes, it is critical to collect various datasets including metadata, visual data, and transient data that cover different boundary and operating conditions. Unfortunately, collecting all dimensions of the data from computational and experimental efforts has not been addressed in this area. Therefore, it is critical to initiate data clusters following standard procedures discussed and defined by the community. This will enable researchers to access major assets (data and developed ML tools) that are stored in cloud environments and add their own to continue the phase-change cluster development.

In addition, technologies are needed to ensure the safe sharing of data since data owners take on risks when sharing their data with the research community. For example, CryptoNets88 allows neural networks to operate over encrypted data, ensuring that data remain confidential because decryption keys are not needed in neural networks.[111] At the same time, privacy methods must remain sufficiently explainable and transparent to help researchers correct them and make them safe, efficient, and accurate.

With the growth of training data and the complexity of deep learning models, scalable algorithms and software become necessities for solving large-scale problems and resolving fine-scale physics. Training a neural network is generally a time-consuming and challenging process, where the difficulty of the training process scales as the network becomes larger and deeper. Moreover, approaches that incorporate PDE-based soft regularization only add to this difficulty by making the loss landscape even harder to optimize. Accordingly, the benefits of the developed DL tools outweigh the high costs and challenges associated with their training only if the resulting models are generalizable and scalable. A promising approach to improving the scalability of trained models inspired by curriculum learning is to first train with simpler constraints and then couple pre-trained models using domain decomposition strategies to solve the target problem.[58] Another complementary approach is to build foundational models for ML models that embrace the "pre-train and fine-tune" paradigm for science problems. This paradigm has been highly successful in other domains and holds great promise in accelerating scientific advancements, enhancing model performance, and facilitating efficient transfer of knowledge across related scientific domains. Building on this progress, creating reusable foundational models to learn unsteady single-phase physics in two-phase processes and coupling these models to make predictions for multi-physics phenomena presents one promising future direction for building a generalizable and sustainable CI.

***Communicating across multiple disciplines.***

This review also elucidates the important role that AI can play in transferring knowledge between multiple disciplines, perhaps with emphasis on material and thermofluidic sciences. The nucleation phenomena described in this paper are highly dependent on surface properties (based on their structures and chemistry). Heat transfer performance can be enhanced using nano- or micro-materials that have desired heat and mass transfer properties. The comprehensive understanding about phase-change physics opens new opportunities for materials scientists in the context of materials design. In earlier studies, these processes were not efficient as scientists needed to search through a large range of materials and architectures, but now they can leverage AI-based design tools. For example, AI-assisted data mining toolkits like matminer have seen success in the material sciences by aiding the development and testing of new materials.[112] With access to new datasets, including metadata, visual data, and transient data, a novel collaboration between the heat transfer and materials science communities becomes feasible. We can formulate topology optimization models, explore inverse design techniques, and delve into the realm of multi-objective design as new dimensions in this research. During this process, building an additional layer of communications with computer and data scientists becomes essential. Many of the AI technologies discussed in this review are prototypes derived from successful models developed by computer scientists. These models can aid in the creation of task-specific new models for two-phase tasks. Data scientists can help develop and optimize data architectures that efficiently curate large data streams into meaningful features and identify important data features to address phase-change heat transfer. Successful studies through this convergence will provide a holistic description of phase-change heat transfer dynamics, enabling blueprints for next-generation phase-change thermal management designs. These new designs will, in turn, be translated to many applications, including energy conversion devices, two-phase electronics cooling devices, and water-energy surfaces.

**References**

1. Mousa, M. H., Yang, C.-M., Nawaz, K. & Miljkovic, N. Review of heat transfer enhancement techniques in two-phase flows for highly efficient and sustainable cooling. *Renewable and Sustainable Energy Reviews*, 111896 (2021).
2. Attinger, D. *et al.* Surface engineering for phase change heat transfer: A review. *MRS Energy & Sustainability* **1** (2014).
3. Griffith, P. & Wallis, J. D. The role of surface conditions in nucleate boiling. (Cambridge, Mass.: Massachusetts Institute of Technology, Division of …, 1958).
4. Bankoff, S. Entrapment of gas in the spreading of a liquid over a rough surface. *AIChE journal* **4**, 24-26 (1958).
5. Mccormick, J. L. & Westwater, J. W. Nucleation Sites for Dropwise Condensation. *Chem Eng Sci* **20**, 1021-+ (1965).
6. McCormick, J. & Westwater, J. Nucleation sites for dropwise condensation. *Chem Eng Sci* **20**, 1021-1036 (1965).
7. Zuber, N. Hydrodynamic aspects of boiling heat transfer. *Doctoral Dissertation of University of California* (1959).
8. Plesset, M. S. & Zwick, S. A. The Growth of Vapor Bubbles in Superheated Liquids. *J Appl Phys* **25**, 493-500 (1954).
9. Tanaka, H. A Theoretical Study of Dropwise Condensation. *Journal of Heat Transfer* **97**, 72-78, doi:10.1115/1.3450291 (1975).
10. Miljkovic, N., Enright, R. & Wang, E. N. Modeling and Optimization of Superhydrophobic Condensation. *J Heat Trans-T Asme* **135** (2013).
11. Ivey, H. Relationships between bubble frequency, departure diameter and rise velocity in nucleate boiling. *International Journal of Heat and Mass Transfer* **10**, 1023-1040 (1967).
12. Dadhich, M. & Prajapati, O. S. A brief review on factors affecting flow and pool boiling. *Renew Sust Energ Rev* **112**, 607-625 (2019).
13. Lee, J. *et al.* Deep Vision-Inspired Bubble Dynamics on Hybrid Nanowires with Dual Wettability. *arXiv preprint arXiv:2202.09417* (2022).
14. Suh, Y. *et al.* A Deep Learning Perspective on Dropwise Condensation. *Adv Sci* **8** (2021).
15. Goertzel, B. & Pennachin, C. *Artificial general intelligence*. Vol. 2 (Springer, 2007).
16. Saha, S. *et al.* Hierarchical Deep Learning Neural Network (HiDeNN): An artificial intelligence (AI) framework for computational science and engineering. *Comput Method Appl M* **373** (2021).
17. Goodfellow, I., Bengio, Y. & Courville, A. *Deep learning*. (MIT press, 2016).
18. Tshitoyan, V. *et al.* Unsupervised word embeddings capture latent knowledge from materials science literature. *Nature* **571**, 95-+ (2019).
19. Jain, A. *et al.* Commentary: The Materials Project: A materials genome approach to accelerating materials innovation. *Apl Mater* **1** (2013).
20. Mozaffar, M. *et al.* Deep learning predicts path-dependent plasticity. *P Natl Acad Sci USA* **116**, 26414-26420 (2019).
21. Bostanabad, R. *et al.* Computational microstructure characterization and reconstruction: Review of the state-of-the-art techniques. *Prog Mater Sci* **95**, 1-41 (2018).
22. Raissi, M., Yazdani, A. & Karniadakis, G. E. Hidden fluid mechanics: Learning velocity and pressure fields from flow visualizations. *Science* **367**, 1026-+ (2020).
23. Hughes, M. T., Kini, G. & Garimella, S. Status, Challenges, and Potential for Machine Learning in Understanding and Applying Heat Transfer Phenomena. *J Heat Trans-T Asme* **143** (2021).
24. Wang, X. L. *et al.* A comprehensive review on the application of nanofluid in heat pipe based on the machine learning: Theory, application and prediction. *Renew Sust Energ Rev* **150** (2021).
25. Wang, Z. Y. *et al.* Advanced big-data/machine-learning techniques for optimization and performance enhancement of the heat pipe technology-A review and prospective study. *Appl Energ* **294** (2021).
26. Ahmadi, M. H., Kumar, R., Assad, M. E. & Ngo, P. T. T. Applications of machine learning methods in modeling various types of heat pipes: a review. *J Therm Anal Calorim* **146**, 2333-2341 (2021).
27. Maleki, A., Haghighi, A. & Mahariq, I. Machine learning-based approaches for modeling thermophysical properties of hybrid nanofluids: A comprehensive review. *J Mol Liq* **322** (2021).

28  Ma, T., Guo, Z. X., Lin, M. & Wang, Q. W. Recent trends on nanofluid heat transfer machine learning research applied to renewable energy. *Renew Sust Energ Rev* **138** (2021).
29  Cho, H. J., Preston, D. J., Zhu, Y. Y. & Wang, E. N. Nanoengineered materials for liquid-vapour phase-change heat transfer. *Nat Rev Mater* **2** (2017).
30  Samek, W., Montavon, G., Lapuschkin, S., Anders, C. J. & Muller, K. R. Explaining Deep Neural Networks and Beyond: A Review of Methods and Applications. *P Ieee* **109**, 247-278 (2021).
31  LeCun, Y., Bengio, Y. & Hinton, G. Deep learning. *Nature* **521**, 436-444 (2015).
32  Rosenblatt, F. The Perceptron - a Probabilistic Model for Information-Storage and Organization in the Brain. *Psychol Rev* **65**, 386-408 (1958).
33  Choi, R. Y., Coyner, A. S., Kalpathy-Cramer, J., Chiang, M. F. & Campbell, J. P. Introduction to Machine Learning, Neural Networks, and Deep Learning. *Transl Vis Sci Techn* **9** (2020).
34  Hosny, A., Parmar, C., Quackenbush, J., Schwartz, L. H. & Aerts, H. J. W. L. Artificial intelligence in radiology. *Nat Rev Cancer* **18**, 500-510 (2018).
35  Mater, A. C. & Coote, M. L. Deep Learning in Chemistry. *J Chem Inf Model* **59**, 2545-2559 (2019).
36  Li, X. H. *et al.* Transfer learning in computer vision tasks: Remember where you come from. *Image Vision Comput* **93** (2020).
37  Deng, J. *et al.* in *2009 IEEE conference on computer vision and pattern recognition.*  248-255 (Ieee).
38  Lin, T.-Y. *et al.* in *European conference on computer vision.*  740-755 (Springer).
39  Pan, S. J. & Yang, Q. A survey on transfer learning. *IEEE Transactions on knowledge and data engineering* **22**, 1345-1359 (2009).
40  Hafiz, A. M. & Bhat, G. M. A survey on instance segmentation: state of the art. *Int J Multimed Inf R* **9**, 171-189 (2020).
41  Zhao, Z. Q., Zheng, P., Xu, S. T. & Wu, X. D. Object Detection With Deep Learning: A Review. *Ieee T Neur Net Lear* **30**, 3212-3232 (2019).
42  Long, J., Shelhamer, E. & Darrell, T. in *Proceedings of the IEEE conference on computer vision and pattern recognition.*  3431-3440.
43  Sun, Y. L., Tang, Y., Zhang, S. W., Yuan, W. & Tang, H. A review on fabrication and pool boiling enhancement of three-dimensional complex structures. *Renew Sust Energ Rev* **162** (2022).
44  Hou, Y. M., Yu, M., Chen, X. M., Wang, Z. K. & Yao, S. H. Recurrent Filmwise and Dropwise Condensation on a Beetle Mimetic Surface. *Acs Nano* **9**, 71-81 (2015).
45  Pham, Q. N. *et al.* Boiling Heat Transfer with a Well-Ordered Microporous Architecture. *Acs Appl Mater Inter* **12**, 19174-19183 (2020).
46  Kim, S. H., Chu, I. C., Choi, M. H. & Euh, D. J. Mechanism study of departure of nucleate boiling on forced convective channel flow boiling. *International Journal of Heat and Mass Transfer* **126**, 1049-1058 (2018).
47  Bard, A., Qiu, Y., Kharangate, C. R. & French, R. Consolidated modeling and prediction of heat transfer coefficients for saturated flow boiling in mini/micro-channels using machine learning methods. *Appl Therm Eng* **210**, 118305 (2022).
48  Neto, M. P. & Paulovich, F. V. Explainable matrix-visualization for global and local interpretability of random forest classification ensembles. *IEEE Transactions on Visualization and Computer Graphics* **27**, 1427-1437 (2020).
49  Yang, Y., Morillo, I. G. & Hospedales, T. M. Deep neural decision trees. *arXiv preprint arXiv:1806.06988* (2018).
50  Roßbach, P. Neural networks vs. random forests–does it always have to be deep learning. *Germany: Frankfurt School of Finance and Management* (2018).
51  Hassanpour, M., Vaferi, B. & Masoumi, M. E. Estimation of pool boiling heat transfer coefficient of alumina water-based nanofluids by various artificial intelligence (AI) approaches. *Appl Therm Eng* **128**, 1208-1222 (2018).
52  Mehrabi, M. & Abadi, S. M. A. N. R. Modeling of condensation heat transfer coefficients and flow regimes in flattened channels. *Int Commun Heat Mass* **126** (2021).
53  Kim, H., Moon, J., Hong, D., Cha, E. & Yun, B. Prediction of critical heat flux for narrow rectangular channels in a steady state condition using machine learning. *Nuclear Engineering and Technology* **53**, 1796-1809 (2021).
54  Lee, D. H. *et al.* Application of the machine learning technique for the development of a condensation heat transfer model for a passive containment cooling system. *Nuclear Engineering and Technology* (2021).

55  Kim, K. M., Hurley, P. & Duarte, J. P. Physics-informed machine learning-aided framework for prediction of minimum film boiling temperature. *International Journal of Heat and Mass Transfer* **191**, 122839 (2022).
56  Zhao, X. G., Shirvan, K., Salko, R. K. & Guo, F. D. On the prediction of critical heat flux using a physics-informed machine learning-aided framework. *Appl Therm Eng* **164** (2020).
57  Karniadakis, G. E. *et al.* Physics-informed machine learning. *Nat Rev Phys* **3**, 422-440 (2021).
58  Wang, H. J., Planas, R., Chandramowlishwaran, A. & Bostanabad, R. Mosaic flows: A transferable deep learning framework for solving PDEs on unseen domains. *Comput Method Appl M* **389** (2022).
59  Cai, S. *et al.* Flow over an espresso cup: inferring 3-D velocity and pressure fields from tomographic background oriented Schlieren via physics-informed neural networks. *Journal of Fluid Mechanics* **915** (2021).
60  Schmidt, E., Schurig, W. & Sellschopp, W. Versuche über die Kondensation von Wasserdampf in Film-und Tropfenform. *Technische Mechanik und Thermodynamik* **1**, 53-63 (1930).
61  Nukiyama, S. The maximum and minimum values of the heat Q transmitted from metal to boiling water under atmospheric pressure. *International Journal of Heat and Mass Transfer* **9**, 1419-1433 (1966).
62  Benjamin, R. & Balakrishnan, A. Nucleate pool boiling heat transfer of pure liquids at low to moderate heat fluxes. *International Journal of Heat and Mass Transfer* **39**, 2495-2504 (1996).
63  Graham, R. W. & Hendricks, R. C. Assessment of convection, conduction, and evaporation in nucleate boiling. (1967).
64  Han, C.-Y. *The mechanism of heat transfer in nucleate pool boiling*, Massachusetts Institute of Technology, (1962).
65  Judd, R. & Hwang, K. A comprehensive model for nucleate pool boiling heat transfer including microlayer evaporation. (1976).
66  Mikic, B. & Rohsenow, W. A new correlation of pool-boiling data including the effect of heating surface characteristics. (1969).
67  Podowski, M. Z., Alajbegovic, A., Kurul, N., Drew, D. & Lahey Jr, R. Mechanistic modeling of CHF in forced-convection subcooled boiling. (Knolls Atomic Power Lab., Schenectady, NY (United States), 1997).
68  Suh, Y., Bostanabad, R. & Won, Y. Deep learning predicts boiling heat transfer. *Sci Rep-Uk* **11** (2021).
69  Jin, Y. & Shirvan, K. Study of the film boiling heat transfer and two-phase flow interface behavior using image processing. *International Journal of Heat and Mass Transfer* **177** (2021).
70  Kulenovic, R., Mertz, R. & Groll, M. in *Thermal Sciences 2000. Proceedings of the International Thermal Science Seminar. Volume 1.*   (Begel House Inc.).
71  Maurus, R., Ilchenko, V. & Sattelmayer, T. Study of the bubble characteristics and the local void fraction in subcooled flow boiling using digital imaging and analysing techniques. *Exp Therm Fluid Sci* **26**, 147-155 (2002).
72  Surtaev, A., Serdyukov, V., Zhou, J. J., Pavlenko, A. & Tumanov, V. An experimental study of vapor bubbles dynamics at water and ethanol pool boiling at low and high heat fluxes. *International Journal of Heat and Mass Transfer* **126**, 297-311 (2018).
73  Watanabe, N. & Aritomi, M. Correlative relationship between geometric arrangement of drops in dropwise condensation and heat transfer coefficient. *International Journal of Heat and Mass Transfer* **105**, 597-609 (2017).
74  Maiti, N., Desai, U. B. & Ray, A. K. Application of mathematical morphology in measurement of droplet size distribution in dropwise condensation. *Thin Solid Films* **376**, 16-25 (2000).
75  Damoulakis, G., Gukeh, M. J., Moitra, S. & Megaridis, C. M. in *2021 20th IEEE Intersociety Conference on Thermal and Thermomechanical Phenomena in Electronic Systems (iTherm).*  1015-1023 (IEEE).
76  Wikramanayake, E. & Bahadur, V. in *Heat Transfer Summer Conference.*  V001T013A001 (American Society of Mechanical Engineers).
77  Castillo, J. E., Weibel, J. A. & Garimella, S. V. The effect of relative humidity on dropwise condensation dynamics. *International Journal of Heat and Mass Transfer* **80**, 759-766 (2015).
78  Castillo, J. E. & Weibel, J. A. Predicting the growth of many droplets during vapor-diffusion-driven dropwise condensation experiments using the point sink superposition method. *International Journal of Heat and Mass Transfer* **133**, 641-651 (2019).

79  Watanabe, N., Aritomi, M. & Machida, A. Time-series characteristics and geometric structures of drop-size distribution density in dropwise condensation. *International Journal of Heat and Mass Transfer* **76**, 467-483 (2014).
80  Parin, R. *et al.* Heat transfer and droplet population during dropwise condensation on durable coatings. *Appl Therm Eng* **179** (2020).
81  O'Mahony, N. *et al.* in *Science and information conference.* 128-144 (Springer).
82  Khodakarami, S., Rabbi, K. F., Suh, Y., Won, Y. & Miljkovic, N. Machine learning enabled condensation heat transfer measurement. *International Journal of Heat and Mass Transfer* **194**, 123016 (2022).
83  Hoang, N. H., Song, C. H., Chu, I. C. & Euh, D. J. A bubble dynamics-based model for wall heat flux partitioning during nucleate flow boiling. *International Journal of Heat and Mass Transfer* **112**, 454-464 (2017).
84  Gerardi, C., Buongiorno, J., Hu, L. W. & McKrell, T. Study of bubble growth in water pool boiling through synchronized, infrared thermometry and high-speed video. *International Journal of Heat and Mass Transfer* **53**, 4185-4192 (2010).
85  Kim, S. H. *et al.* Heat flux partitioning analysis of pool boiling on micro structured surface using infrared visualization. *International Journal of Heat and Mass Transfer* **102**, 756-765 (2016).
86  Milan, A., Leal-Taixé, L., Reid, I., Roth, S. & Schindler, K. MOT16: A benchmark for multi-object tracking. *arXiv preprint arXiv:1603.00831* (2016).
87  Moen, E. *et al.* Deep learning for cellular image analysis. *Nat Methods* **16**, 1233-1246 (2019).
88  Suh, Y., Chang, S. H., Simadiris, P., Inouye, T. B., Hoque, M. J., Khodakarami, S., Kharangate, C., Miljkovic, N. & Won, Y. VISION-iT: Deep nuclei tracking framework for digitalizing bubbles and droplets. *SSRN: https://papers.ssrn.com/sol3/papers.cfm?abstract_id=4491956 (2023)*.
89  Newby, J. M., Schaefer, A. M., Lee, P. T., Forest, M. G. & Lai, S. K. Convolutional neural networks automate detection for tracking of submicron-scale particles in 2D and 3D. *P Natl Acad Sci USA* **115**, 9026-9031 (2018).
90  Kang, S. & Cho, K. Conditional Molecular Design with Deep Generative Models. *J Chem Inf Model* **59**, 43-52 (2019).
91  Denton, E. & Fergus, R. in *International conference on machine learning.* 1174-1183 (PMLR).
92  Hessenkemper, H., Starke, S., Atassi, Y., Ziegenhein, T. & Lucas, D. Bubble identification from images with machine learning methods. *arXiv preprint arXiv:2202.03107* (2022).
93  Li, J. Q., Shao, S. Y. & Hong, J. R. Machine learning shadowgraph for particle size and shape characterization. *Meas Sci Technol* **32** (2021).
94  Cerqueira, R. F. L. & Sinmec, E. E. P. Development of a deep learning-based image processing technique for bubble pattern recognition and shape reconstruction in dense bubbly flows. *Chem Eng Sci* **230** (2021).
95  Torisaki, S. & Miwa, S. Robust bubble feature extraction in gas-liquid two-phase flow using object detection technique. *J Nucl Sci Technol* **57**, 1231-1244 (2020).
96  Ma, C. *et al.* Condensation Droplet Sieve. *arXiv preprint arXiv:2202.02940* (2022).
97  Yan, J. Y., Ma, R. & Du, X. Consistent optical surface inspection based on open environment droplet size-controlled condensation figures. *Meas Sci Technol* **32** (2021).
98  Cheng, L. in *Microchannel Phase Change Transport Phenomena* 141-191 (Elsevier, 2016).
99  Hobold, G. M. & da Silva, A. K. Machine learning classification of boiling regimes with low speed, direct and indirect visualization. *International Journal of Heat and Mass Transfer* **125**, 1296-1309 (2018).
100  Mishkinis, D. & Ochterbeck, J. Homogeneous Nucleation and the Heat-Pipe Boiling Limitation. *Journal of engineering physics and thermophysics* **76**, 813-818 (2003).
101  Nie, F. *et al.* Image identification for two-phase flow patterns based on CNN algorithms. *Int J Multiphas Flow* **152** (2022).
102  Hobold, G. M. & da Silva, A. K. Visualization-based nucleate boiling heat flux quantification using machine learning. *International Journal of Heat and Mass Transfer* **134**, 511-520 (2019).
103  Xu, H., Tang, T., Zhang, B. R. & Liu, Y. C. Identification of two-phase flow regime in the energy industry based on modified convolutional neural network. *Prog Nucl Energ* **147** (2022).
104  Ambrosio, J. D., Lazzaretti, A. E., Pipa, D. R. & da Silva, M. J. Two-phase flow pattern classification based on void fraction time series and machine learning. *Flow Meas Instrum* **83** (2022).

105    Maulud, D. & Abdulazeez, A. M. A review on linear regression comprehensive in machine learning. *Journal of Applied Science and Technology Trends* **1**, 140-147 (2020).
106    Lotfian, M., Ingensand, J. & Brovelli, M. A. The Partnership of Citizen Science and Machine Learning: Benefits, Risks, and Future Challenges for Engagement, Data Collection, and Data Quality. *Sustainability-Basel* **13** (2021).
107    Rokoni, A. *et al.* Learning new physical descriptors from reduced-order analysis of bubble dynamics in boiling heat transfer. *International Journal of Heat and Mass Transfer* **186**, 122501 (2022).
108    Gallego, G. *et al.* Event-based vision: A survey. *IEEE transactions on pattern analysis and machine intelligence* **44**, 154-180 (2020).
109    Lu, D. Y., Suh, Y. & Won, Y. in *micro Flow and Interfacial Phenomena Conference (microFip).*
110    Sironi, A., Brambilla, M., Bourdis, N., Lagorce, X. & Benosman, R. in *Proceedings of the IEEE Conference on Computer Vision and Pattern Recognition.* 1731-1740.
111    Gilad-Bachrach, R. *et al.* in *International conference on machine learning.* 201-210 (PMLR).
112    Ward, L., Dunn, A., Faghaninia, A., Zimmermann, N. E., Bajaj, S., Wang, Q., Montoya, J., Chen, J., Bystrom, K., Dyalla, M., Asta, M., Persson, K. A., Snyder, G. J., Foster, I. & Jain, A. Matminer: An open source toolkit for materials data mining. *Computational Materials Science*, **152**, 60-69 (2018).
113    Upot, N. V., Rabbi, K. F., Khodakarami, S., Ho J. Y., Mendizabal, J. K. & Miljkovic, N. Advances in micro and nanoengineered surfaces for enhanced boiling and condensation heat transfer: a review. *Nanoscale Advances* **5**, 1232-1270 (2023).
114    Baker, N., Alexander, F., Bremer, T., Hagberg, A., Kevrekidis, Y., Najm, H., Parashar, M., Patra, A., Sethian, J., Wild, S. and Willcox, K. *Workshop report on basic research needs for scientific machine learning: Core technologies for artificial intelligence*. USDOE Office of Science (SC), Washington, DC (United States) (2019).
115    Lu, L., Jin, P., Pang, G., Zhang, Z. and Karniadakis, G.E. Learning nonlinear operators via DeepONet based on the universal approximation theorem of operators. *Nature machine intelligence*, **3**, pp.218-229 (2021).
116    Li, Z., Kovachki, N., Azizzadenesheli, K., Liu, B., Bhattacharya, K., Stuart, A. and Anandkumar, A. Fourier neural operator for parametric partial differential equations. *arXiv preprint arXiv:2010.08895* (2020).
117    Hassan, S. M. S., Feeney, A., Dhruv, A., Kim, J., Suh, Y., Ryu, J., Won, Y. & Chandramowlishwaran, A. BubbleML: A Multi-Physics Dataset and Benchmarks for Machine Learning (1.0). *Zenodo.* https://doi.org/10.5281/zenodo.8039787 (2023).
118    Seong, J.H., Ravichandran, M., Su, G., Phillips, B. & Bucci, M. Automated bubble analysis of high-speed subcooled flow boiling images using U-net transfer learning and global optical flow. *International Journal of Multiphase Flow* **159**, 104336 (2023)
119    Chang, S., Suh, Y., Shingote, C., Huang, C. N., Mudawar, I., Kharangate, C. & Won, Y. Autonomous visualization of digital flow bubbles for predicting critical heat flux *SSRN* 4458770 (2023)
120    Li, Z., Zheng, H., Kovachki, N., Jin, D., Chen, H., Liu, B., Azizzadenesheli, K. and Anandkumar, A. Physics-informed neural operator for learning partial differential equations. *arXiv preprint arXiv:2111.03794*. (2021)

**Acknowledgments**

The authors give thanks to Tiffany B. Inouye for assistance in the preparation of this manuscript. The authors gratefully acknowledge the funding support from the Office of Naval Research (ONR) with Dr. Mark Spector as the program manager (Grant No. N00014-22-1-2063) and the National Science Foundation under Award No. TTP 2045322.

**Author information**
Authors and Affiliations
**Department of Mechanical and Aerospace Engineering, University of California Irvine, CA, USA**
Youngjoon Suh, Aparna Chandramowlishwaran, Yoonjin Won

**Department of Engineering and Computer Science, University of California, Irvine, CA, USA**
Aparna Chandramowlishwaran, Yoonjin Won

Contributions

Y.S., A.C., and Y.W. conceptualized the article structure, content, and figures. Y.S., A.C., and Y.W. wrote and edited the manuscript.

Corresponding author

Correspondence to Yoonjin Won.

**Ethics declarations**

Competing interests

The authors declare no competing interests.